\documentclass[fleqn,10pt]{wlscirep}
\usepackage[T1]{fontenc}

\usepackage{threeparttable}
\usepackage{booktabs}
\usepackage{siunitx}
\usepackage{array}
\usepackage{geometry}
\usepackage{amsmath} 

\usepackage{graphicx}
\usepackage{float}
\usepackage{rotating}
\usepackage{longtable}
\usepackage{dcolumn}
\usepackage{bm}
\usepackage{color}
\usepackage{xcolor}

\usepackage[caption=false]{subfig}

\usepackage{dcolumn}
\usepackage{bm}
\usepackage[normalem]{ulem}
\usepackage{multirow}

\newcommand{\wideunderline}[2][2em]{%
	\underline{\makebox[\ifdim\width>#1\width\else#1\fi]{#2}}%
}

\newcommand{\be}{\begin{equation}}
\newcommand{\ee}{\end{equation}}
\newcommand{\bea}{\begin{eqnarray}}
\newcommand{\eea}{\end{eqnarray}}

\title{Two-dimensional M$_2$X$_2$(M=transition-metal; X=S, Se, Te) family with emerging semiconducting, semimetallic, and magnetic properties}
\author[1]{Y. Yekta}

\author[1]{H. R. Ramezani}
\affil[1]{Department of Physics, University of Guilan, 41335-1914, Rasht, Iran} 
\author[1]{H. Hadipour}
\affil{2nd Institute of Physics C, RWTH Aachen University, 52074 Aachen, Germany}

\author[2,$\dagger$]{S. A. Jafari}
\affil[$\dagger$]{akbar.jafari@rwth-aachen.de}

	\begin{abstract}
	
	The exploration for novel two-dimensional (2D) materials with diverse electronic characteristics has
	attracted growing interest in recent years. Using density functional theory (DFT) calculations, we have predicted a new family of 2D transition-metal (TM) based compounds under the nomenclature M$_2$X$_2$ (where M represents TMs, and X denotes chalcogen elements like S, Se, and Te). 
	Our investigation delves into the examination of the formation energies, dynamical/thermal stabilities, mechanical properties, electronic structures, and magnetic properties of various systems within this family. Through our computational analyses, we have discovered a total of 35 thermodynamically and dynamically stable M$_2$X$_2$ monolayer materials that exhibit remarkable diversity in terms of their electronic and magnetic properties. Our findings will pave the way for the experimental realization of various M$_2$X$_2$ structures in the near future.
	In particular, among the predicted compounds, M$_2$X$_2$(M=Zn, Cd; X=S, Se, Te) are a direct
	band-gap semiconductor with band gaps between 0.9 to 2.6 eV (1.3 to 3.7 eV) by DFT+PBE (hybrid functional HSE) calculations.
	M$_2$X$_2$(M=Ti, Zr, Hf, Tc, Re) are zero-gap
	semiconductor (semimetals) in standard DFT+PBE calculation. Inclusion of spin–orbit
	coupling leads to a gap opening of 0.1 eV. 
	Notably, our analysis has also unveiled the magnetic nature of certain materials, such as Mn$_2$X$_2$(X=S, Se), Fe$_2$X$_2$(X=Se, Te), and Ti$_2$Te$_2$. 
	The prediction of semiconducting (magnetic) M$_2$X$_2$ materials not only offers valuable insights into the underlying electronic properties (magnetism) of 2D systems but also positions these materials as promising candidates for the development of advanced electronic (spintronic) devices.
	
\end{abstract}

\begin{document}


	\date{\today}


	
\maketitle

\section{Introduction}\label{sec1}

Two-dimensional (2D) materials have gained significant attention in recent years due to their unique electronic~\cite{Ahn,Li-MoS2}, magnetic~\cite{Avsar-review-mag}, and optical~\cite{Suk} properties.
When these 2D materials host transition-metal (TM) atoms with $\textnormal d$ electrons,  they continue to furnish richer physical properties driven by their electronic structures and spin states, as well as the significant spin-orbit coupling (SOC) of the TMs.  
The interplay between the internal degrees of freedom of electrons, including charge, orbital, and spin holds great potential for both fundamental research and practical device applications. The literature contains numerous examples of TM-based low-dimensional systems, which have been or may potentially be exfoliated in experimental studies.
Some important examples of these 2D materials are TM dichalcogenides (TMDs)~\cite{Chen,Manzeli,Ramezani} like MoS$_2$, MXenes~\cite{Gogotsi,Yekta} such as Mo$_2$C, TM halides~\cite{Huang} such as CrI$_3$, and newly synthesized MA$_2$Z$_4$ family (M=elements of TM; A=Si, Ge; Z=N, P, As)~\cite{Hong,Wang-1,Yin}.

In this paper we use density functional theory (DFT) calculations to predict new 2D TM-based family M$_2$X$_2$ (M=TM; X=S, Se, Te).  The side and top view crystal structure of the monolayers M$_2$X$_2$ are depicted in Figs.\,\ref{fig1-Cry}(a) and \,\ref{fig1-Cry}(b). These materials can be viewed as a AB stacking of two honeycomb MX layers. In AB stacking, the layers are arranged such that one layer is directly on top of the other, with the B atoms of the second layer sitting directly on top of the A atoms of the first layer. Each MX layer has a structure similar to graphene but consists of alternating M (sublattice A) and X (sublattice B) atoms arranged in a buckled hexagonal lattice.
We find 35, thermodynamically and dynamically, stable M$_2$X$_2$ monolayer materials,
which exhibit diverse electronic and magnetic properties. Therefore some of the M$_2$X$_2$ materials are expected to be realized in the experiment.
Our first-principles calculations indicate that the most stable M$_2$X$_2$ systems are metallic. 
The states around the Fermi energy ($E_\textnormal F$) of these compounds are primarily dominated by the $\textnormal d$ orbitals of the TM.
Among the predicted compounds, M$_2$X$_2$ (where M=Zn, Cd and X=S, Se, Te) are identified as direct band-gap semiconductors, exhibiting band gaps $E_\textnormal g$ ranging from 0.9 to 2.6 eV as determined by Perdew-Burke-Ernzerhof (PBE)~\cite{Perdew} calculations. 
In contrast, the compounds M$_2$X$_2$ (with M=Ti, Zr, Hf, Tc, Re) behave as zero-gap semiconductors or semimetals when analyzed using standard DFT+PBE calculations.
However, the inclusion of SOC modifies their  band structure, resulting in the opening of a band gap of approximately $0.1$ eV.
This broad range of band gaps suggests a  potential for diverse applications in optoelectronic devices, particularly in areas requiring tunable electronic properties.
Further, the calculation is continued using the computationally more expensive functional of Heyd-Scuseria-Ernzerhof (HSE)~\cite{Heyd} to improve the electronic band structures.
Overall, the findings highlight rich electronic landscape of these materials and their potential for innovative applications in various fields, including photovoltaics, thermoelectrics, and quantum computing.

Due to the the weak covalent bonding between
 the TM and the X element in M$_2$X$_2$,  the number of magnetic systems in the M$_2$X$_2$ family is more than in other 2D TM-based families. In fact, the number of magnetic systems in 2D systems are quite scarce. 
 Experimental observations have identified only a few 2D materials, such as CrX$_3$ (X=Br, I)~\cite{Seyler-CrI3}, Cr$_2$Ge$_2$Te$_6$~\cite{Gong-Cr2Ge2Te6}, and MX$_2$ (M=V, Mn; X=Se, Te)~\cite{Bonilla,Karbala} exhibiting magnetic ordering. Among stable M$_2$X$_2$ materials, there are two antiferromagnetic Mn$_2$X$_2$ (X=S, Se) systems, and three ferromagnetic systems, namely Fe$_2$X$_2$(X=Se, Te), and Ti$_2$Se$_2$. The possibility of magnetic states in M$_2$X$_2$ materials not only provides valuable insights into the underlying magnetism of 2D systems but also qualifies these materials as promising candidates for the development of advanced spintronic devices. By exploring the properties and potential applications of the present new family of materials, one can hope to deepen  comprehension of magnetism at the nanoscale and unlock new possibilities for innovative technologies.

The rest of the paper is organized as follows.
The computational method and crystal structure are presented in Sec.\,\ref{sec2}. Sec.\,\ref{sec3} focuses on the results and discussion, providing a detailed analysis of the dynamical/thermal stability, electronic structure, and magnetic properties of M$_2$X$_2$ monolayers. Finally, in Sec.\,\ref{sec4}, the paper is summarized.

	\section{Crystal structure and symmetry}\label{sec2}

\begin{figure}[!h]
\begin{center}
\includegraphics[scale=0.328]{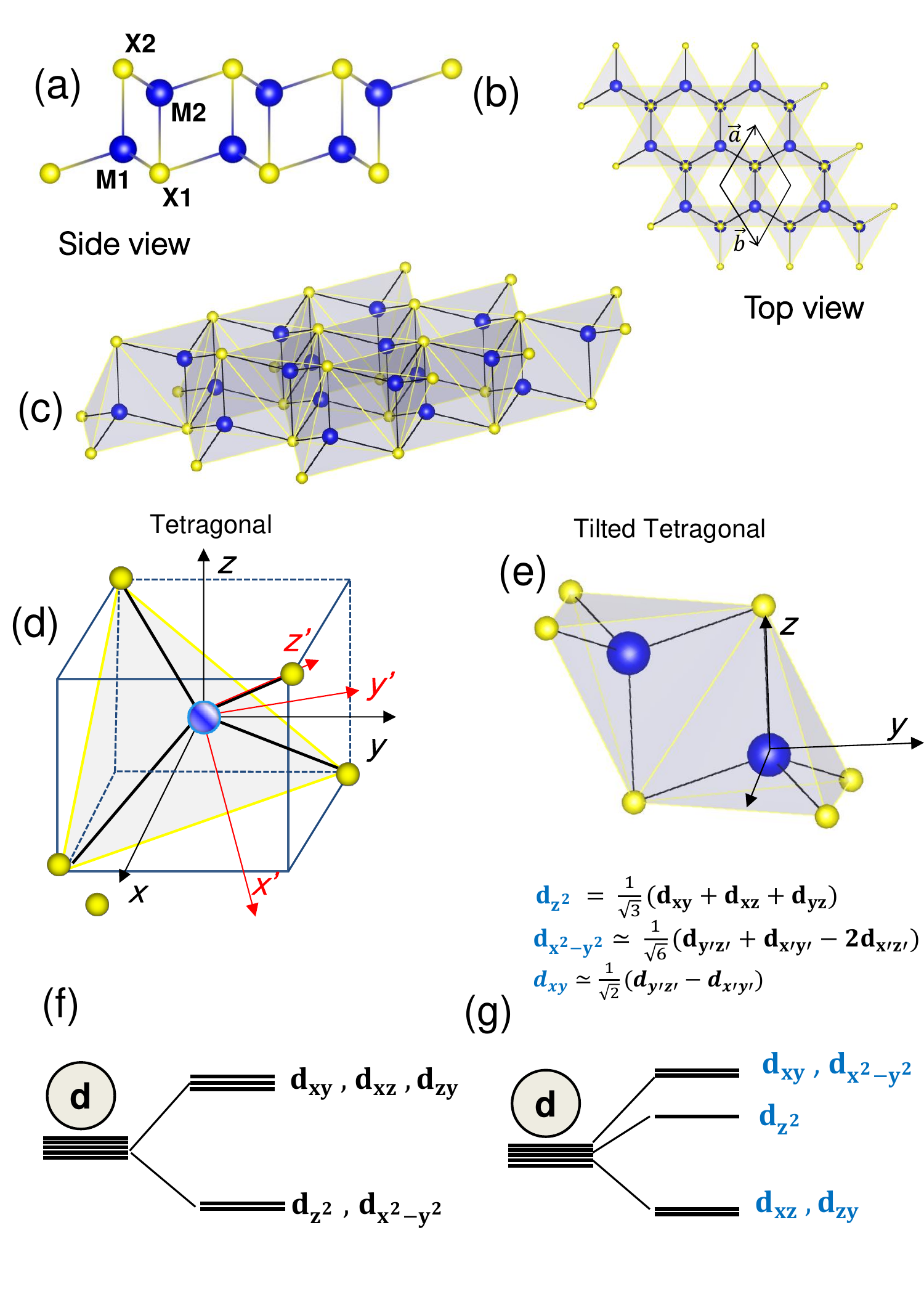}
\end{center}
\vspace*{-0.5cm} 
\caption{(Colors online) (a) Side view and (b) top view of the single layer crystal of M$_2$X$_2$. (c) Presentation of tetrahedrons in side view crystal structure.  (d) 
One TM bonded with four X atoms in conventional tetrahedral coordination.
(e) Tilted tetrahedral coordination of one TM
and four chalcogen X atoms. The blue and yellow spheres denote M and X atoms, respectively. Crystal field splitting of d orbitals in (f) Conventional tetrahedral structure and (g) Tilted tetrahedral structure.} 
\label{fig1-Cry}
\end{figure}

We consider 2D systems with the chemical formula M$_2$X$_2$. Here, M represents TM elements and X are chalcogen elements,
namely S, Se, and Te. 
 The side and top view crystal structure of the monolayers M$_2$X$_2$ are depicted in Figs.\,\ref{fig1-Cry}(a) and \,\ref{fig1-Cry}(b). 
The optimized structural parameters including in-plane lattice
constant ($a$ = $b$), bond lengths between M-X ($\textnormal d^{\mathstrut}_{\textnormal M\mbox{-}\textnormal X}$), and the $z$ component of the atomic positions are given in Table\,\ref{table:struct}. 
These materials with P-3m1 space group can be considered as an AB stacking of two honeycomb MX layers. In this arrangement, one layer is positioned directly above the other, with the B atoms of the second layer aligned directly over the A atoms of the first layer and vice vesa. Each MX layer has a structure akin to graphene, featuring alternating M (sublattice A) and X (sublattice B) atoms organized in a buckled hexagonal lattice.
 
In a M$_2$X$_2$ structure, the central TM atom is surrounded by four ligands X arranged at the corners of a tetrahedron as shown in the Figs.\,\ref{fig1-Cry}(c).
This is completely different from other well-known TM-based compounds like TMDs and MXenes in which, the TM atoms are each bound to six halogen atoms in octahedral or trigonal prismatic coordination~\cite{Manzeli,Gogotsi}. 
Selecting every other vertex of a cube, such that any pair is joined by a diagonal of the cube's face, results in a standard tetrahedron [see Fig.\,\ref{fig1-Cry}(d)]. If we take the
three M-X bonds in M$_2$X$_2$, whose ligand ends form the corners
of one of these tetrahedron faces. As depicted in Fig.\,\ref{fig1-Cry}(d),  in conventional tetrahedral structure, 3-fold rotation axes is in the $z'=\frac{1}{\sqrt3}(i+j+k)$ direction. 
Other two axis $x'$ and $y'$ are located in a plane constructed by three X atoms.

Here in M$_2$X$_2$, the tetrahedron is not like the conventional perovskite
tetrahedral structure. The tetrahedron is tilted with
respect to the standard Cartesian coordinate $x$, $y$, $z$, in
such a way that the 3-fold rotation axes is in the $z$ direction. 
As shown in Fig.\,\ref{fig1-Cry}(e), in fact, one of the four triangular sides
of tetrahedron is lying on the floor to which the $z$ axis is perpendicular.

Two coordinates are connected to each other by matrix: 
\begin{equation}
\begin{pmatrix}

\frac{1}{\sqrt6}  & \frac{1}{\sqrt2} & \frac{1}{\sqrt3}   \\

\frac{1}{\sqrt6} & -\frac{1}{\sqrt2}  & \frac{1}{\sqrt3}  \\

-\frac{2}{\sqrt6}  & 0 &  \frac{1}{\sqrt3}   \\

\end{pmatrix}
\end{equation}
\\

 The effect of this tilting on
the configuration of electrons in the present structure is
discussed in detail in the following.
In standard tetrahedral structure, crystal field splits $\textnormal d$ orbitals into the same $t_{\textnormal{2g}}$ and $e_\textnormal g$ sets of orbitals as
the two orbitals in the $e_\textnormal g$ set are lower in energy than the three orbitals in the $t_{\textnormal{2g}}$ set [see Fig.\,\ref{fig1-Cry}(f)]. 
For example, the orbital orientation in $\textnormal d_{z^2}$ state reduces the energy due to weak overlapping of $\textnormal d$
states with X atoms and weaker Coulomb interaction. 
The situation is different in the tilted tetrahedral structure. 
The tetrahedron is tilted in such a way that $\textnormal d_{z^2}$ orbital has strong overlap with X atoms. 
Therefore, as depicted in Fig.\,\ref{fig1-Cry}(g), the $\textnormal d_{z^2}$ orbitals will be a part of high-energy levels. 
In fact, if we linearly combine the local $t_{2g}$ orbitals as $\frac{1}{\sqrt3}(\textnormal d_{x'y'} + \textnormal d_{y'z'} + \textnormal d_{x'z'})$, we obtain a $\textnormal d_{z^2}$ orbital oriented perpendicular to the layer.
The next two higher energy orbitals are constructed as other combination of $t_{2g}$ states, namely $\textnormal d_{x^2-y^2} \simeq \frac{1}{\sqrt2}(\textnormal d_{y'z'}-\textnormal d_{x'y'})$ and $\textnormal d_{xy} \simeq \frac{1}{\sqrt6}(\textnormal d_{y'z'}+\textnormal d_{x'y'}-2\textnormal d_{x'z'})$. 
Note that the higher energy bands are not of pure $t_{\textnormal{2g}}$ character but exhibit admixture of $e_\textnormal g$ states. 
The described denominations  refer to their dominant orbital character. 
$\textnormal d_{xz}$ and $\textnormal d_{yz}$ are farther from the ligands than the others and therefore experiences less repulsion.

\begin{figure}[!h]
\begin{center}
\includegraphics[scale=0.58]{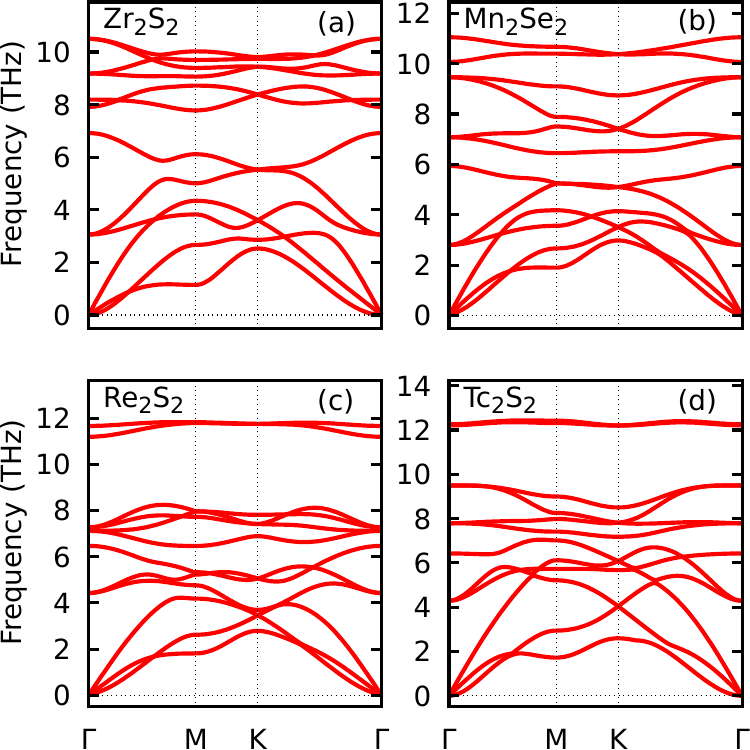}
\end{center}
\vspace*{-0.5cm} 
\caption{(Colors online) Phonon dispersion of M$_2$X$_2$ monolayers (a) Zr$_2$S$_2$, (b) Mn$_2$Se$_2$, (c) Re$_2$S$_2$, and (d) Tc$_2$S$_2$.} 
\label{fig-phonon}
\end{figure}

\begin{figure*}[!]
\begin{center}
\includegraphics[scale=0.72]{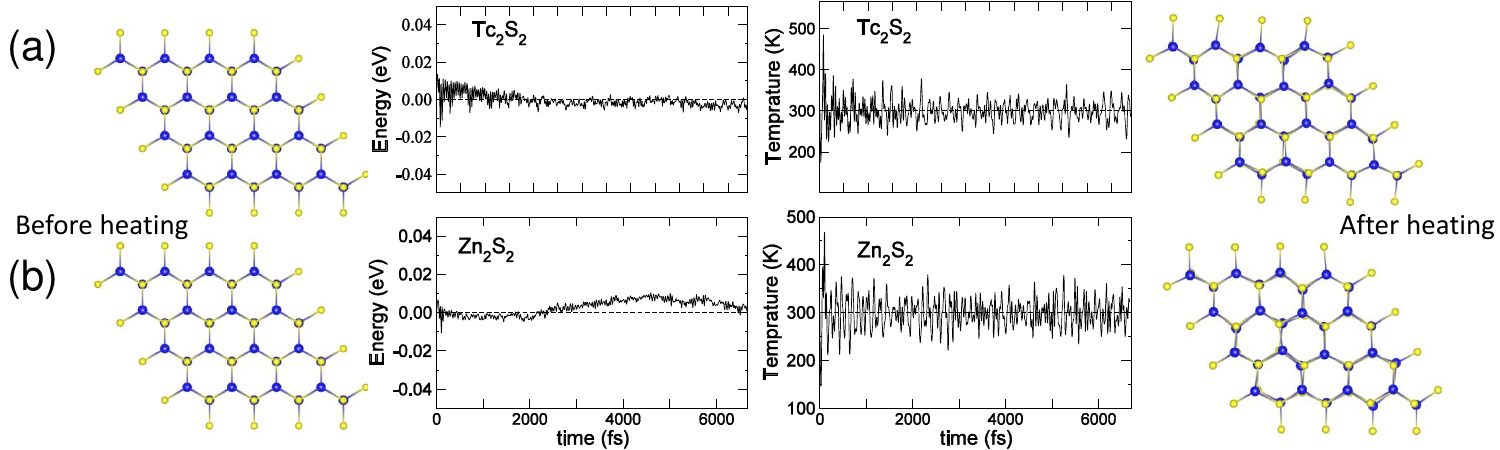}
\end{center}
\vspace*{-0.5cm} 
\caption{(Colors online) (a) and (b) The evolution of energy during the simulation, time-dependent temperature fluctuation, and the final snapshots of the resulting geometries at 300 K of Tc$_2$S$_2$ and Zn$_2$S$_2$ through AIMD calculations.} 
\label{fig-AIMD}
\end{figure*}

\begin{table*}[!ht]
				\sisetup{
		table-format = 1.3,
		table-align-text-pre = false,
		table-align-text-post = false
	}
	\centering
	\begin{threeparttable}
		\caption{Structural and electronic properties of 60 M$_2$X$_2$ candidates.  $a$ is the in-plane
lattice constant. The $x$ and $y$ coordinates of the atomic positions in the order of M1, M2, X1, and X2 are (1/3, 2/3), (2/3, 1/3), (1/3, 2/3), and (2/3, 1/3) respectively in fractional coordinates. $z^{\mathstrut}_{\textnormal{M1}}$, $z^{\mathstrut}_{\textnormal{M2}}$, $z^{\mathstrut}_{\textnormal{X1}}$, and $z^{\mathstrut}_{\textnormal{X2}}$ are the $z$ component of the atomic positions. Bond lengths are given by $\textnormal d^{\mathstrut}_{\textnormal M\mbox{-}\textnormal X}$. $E_\textnormal g$ denote band gap calculated by PBE, PBE+SOC, and HSE06+SOC calculations, respectively.  $E_\textnormal f$ is formation energy.} 
	\scriptsize
			\begin{tabular}{ccccccccccccc}
				\toprule
				No. &M$_2$X$_2$  & $a$(\AA)\tnote{*} & $z^{\mathstrut}_{\textnormal{M1}}$(\AA) & $z^{\mathstrut}_{\textnormal{M2}}$(\AA) & $z^{\mathstrut}_{\textnormal{X1}}$(\AA) & $z^{\mathstrut}_{\textnormal{X2}}$(\AA) &  $\textnormal d^{\mathstrut}_{\textnormal{M1}\mbox{-}\textnormal{X1}}$(\AA)\tnote{$\dagger$} & $\textnormal d^{\mathstrut}_{\textnormal{M1}\mbox{-}\textnormal{X2}}$(\AA)\tnote{$\ddagger$}  & Ground State & $E^{\textnormal{PBE}}_{\textnormal g}$$|$ $E^{\textnormal{SOC}}_{\textnormal g}$ $|$ $E^{\textnormal{HSE06}}_{\textnormal g}$ &
				$E^{\mathstrut}_{\textnormal f}$(eV/f.u) \\
\\[-2.1ex]
\midrule  \addlinespace

1 & $ \rm{Ti_2S_2} $ & 3.92 & 10.30 & 8.56  & 7.82 & 11.04 & 2.48 & 2.38 & Semiconductor & 0.01 $|$ 0.01 $|$  0.99 & -1.12  \\ 
2 & $ \rm{V_2S_2} $ & 3.68 & 10.16 & 8.70  & 7.78 & 11.08 & 2.38 & 2.32 & Metal & -----  & -0.78  \\ 
3 & $ \rm{Cr_2S_2} $ & 3.57 & 10.04 & 8.81  & 7.78 & 11.07 & 2.26 & 2.30 & Magnetic-Metal &  ----- & -0.40    \\ 
4 & $ \rm{Mn_2S_2} $ & 3.58 & 10.02 & 8.84  & 7.89 & 10.97 & 2.13 & 2.27 & Magnetic-Metal & 0.00 $|$  0.00 $|$  0.81 & -0.43    \\ 
5 & $ \rm{Fe_2S_2} $ & 3.56 & 10.20 & 8.65  & 8.03 & 10.82 & 2.17 & 2.15 & Metal & -----  & -0.38    \\ 
6 & $ \rm{Co_2S_2} $ & 3.57 & 10.20 & 8.65  & 8.02 & 10.83 & 2.18 & 2.15 & Metal & -----  & -0.43    \\ 
7 & $ \rm{Ni_2S_2} $ & 3.61 & 10.19& 8.67  & 7.94 &10.92 & 2,21 & 2.25 & Metal & -----  & -0.34    \\ 
8 & $ \rm{Cu_2S_2} $ & 3.79 & 10.33 & 8.52  & 7.96 & 10.89 & 2.37 & 2.26 & Metal & -----  & -0.10    \\ 
9 & $ \rm{Zn_2S_2} $ & 3.92 & 10.45 & 8.41  &  7.87 & 10.99 & 2.33 & 2.58 & Semiconductor & 2.62 $|$  2.56 $|$  3.77  & -0.69  \\ 
10 & $ \rm{Zr_2S_2} $ & 4.20 & 10.39 & 8.47  & 7.76 & 11.10 & 2.63 & 2.53 & Semiconductor & 0.02 $|$  0.11 $|$  0.49 & -1.13  \\
11 & $ \rm{Nb_2S_2} $ & 3.96 & 10.23 & 8.62  & 7.70 & 11.16 & 2.54 & 2.46 & Metal &  ----- & -0.70  \\ 
12 & $ \rm{Mo_2S_2} $ & 3.86 & 10.09 & 8.77  & 7.69 & 11.16 & 2.40 & 2.48 & Magnetic-Metal & -----  & -0.23   \\ 
13 & $ \rm{Tc_2S_2} $ & 3.87 & 10.03 & 8.83  & 7.78 & 11.07 & 2.25 & 2.46 & Semimetal & 0.00 $|$  0.00 $|$  0.63  & -0.23   \\ 
14 & $ \rm{Ru_2S_2} $ & 3.80 & 10.28 & 8.58  & 7.96 & 10.90 & 2.32 & 2.28 & Metal & -----  & -0.15  \\ 
15 & $ \rm{Cd_2S_2} $ & 4.31 & 10.62 & 8.23  & 7.81& 11.05 & 2.53 & 2.81 & Semiconductor & 1.81 $|$  1.79 $|$  2.70  & -0.53   \\ 
16 & $ \rm{Hf_2S_2} $ & 4.13 & 10.37 & 8.49  & 7.78 & 11.08 & 2.59 & 2.49 & Semiconductor & 0.01 $|$  0.10 $|$  0.54  & -0.89  \\ 
17 & $ \rm{Ta_2S_2} $ & 3.93 & 10.23 & 8.63  & 7.69 & 11.17 & 2.54 & 2.45 & Metal & -----  & -0.48  \\ 
18 & $ \rm{W_2S_2} $ & 3.88 & 10.05 & 8.80  & 7.64 & 11.21 & 2.41 & 2.52 & Metal & -----  & 0.07   \\ 
19& $ \rm{Re_2S_2} $ & 3.88 & 10.00 & 8.86  & 7.71 & 11.15 & 2.29 & 2.52 & Semimetal & 0.02 $|$  0.00 $|$  1.06 & 0.10    \\ 
20 & $ \rm{Ir_2S_2} $ & 3.28 & 10.50 & 8.36  &7.10 &11.75 & 2.27 & 3.39 & Semiconductor &  0.03 $|$  0.07 $|$  0.22  & -0.19  \\ \hline
\addlinespace
21 & $ \rm{Ti_2Se_2} $ & 4.05 & 10.28 & 8.57  & 7.65 & 11.21 & 2.63 & 2.51 & Semiconductor &  0.01 $|$  0.04 $|$  0.96  & -0.82   \\ 
22 & $ \rm{V_2Se_2} $ & 3.79 & 10.13 & 8.72  & 7.61 & 11.25 & 2.52 & 2.45 & Metal & -----  & -0.53   \\ 
23 & $ \rm{Cr_2Se_2} $ & 3.65 & 9.98 & 8.87  & 7.59 & 11.27 & 2.40 & 2.47 & Magnetic-Metal & -----  & -0.21    \\ 
24 & $ \rm{Mn_2Se_2} $ & 3.66 & 9.97 & 8.89  & 7.69 & 11.17 & 2.28 & 2.43 & Magnetic-Metal & 0.00 $|$  0.00 $|$  0.68  & -0.33   \\ 
25 & $ \rm{Fe_2Se_2} $ & 3.73 & 10.15 & 8.70  & 7.86 & 10.99 & 2.29 & 2.31 & Magnetic-Metal & -----  & -0.16   \\ 
26 & $ \rm{Co_2Se_2} $ & 3.74 & 10.19 & 8.66  & 7.89 & 10.97 & 2.31 & 2.29 & Metal & -----  & -0.24    \\
27 & $ \rm{Ni_2Se_2} $ & 3.74 & 10.16 & 8.69  & 7.80 & 11.06 & 2.33 & 2.37 & Metal & -----  & -0.26    \\   
28 & $ \rm{Cu_2Se_2} $ & 3.76 & 10.15 & 8.71  & 7.64 & 11.21 & 2.50 & 2.42 & Metal &  ----- & -0.08  \\ 
29 & $ \rm{Zn_2Se_2} $ & 4.08 & 10.38 & 8.48  & 7.72 & 11.14 & 2.47 & 2.66 & Semiconductor & 1.83 $|$ 1.62 $|$  2.63  & -0.61  \\ 
30 & $ \rm{Zr_2Se_2} $ & 4.33 & 10.37 & 8.48  & 7.59 & 11.26 & 2.78 & 2.66 & Semiconductor & 0.00 $|$  0.05 $|$  0.62   & -0.91  \\
31 & $ \rm{Nb_2Se_2} $ & 4.06 & 10.21 & 8.65  & 7.53 & 11.33 & 2.68 & 2.60 & Metal & -----  & -0.49    \\ 
32 & $ \rm{Mo_2Se_2} $ & 3.94 & 10.04 & 8.81  & 7.50 & 11.35 & 2.54 & 2.63 & Metal &  ----- & -0.11    \\
33 & $ \rm{Tc_2Se_2} $ & 3.93 & 10.00 & 8.86  & 7.59 & 11.26 & 2.40 & 2.60 & Semimetal &  0.01 $|$  0.00 $|$  0.82 & -0.10   \\  
34 & $ \rm{Ru_2Se_2} $ & 3.97 & 10.23 & 8.63  & 7.81 & 11.05 & 2.42 & 2.43 & Metal & -----  & 0.03    \\ 
35 & $ \rm{Cd_2Se_2} $ & 4.45 & 10.53 & 8.33  & 7.63 & 11.22 & 2.66 & 2.89 & Semiconductor & 	1.54 $|$  1.40 $|$  2.19  & -0.52    \\ 
36 & $ \rm{Hf_2Se_2} $ & 4.26 & 10.35 & 8.51  & 7.60 & 11.26 & 2.75 & 2.62 & Semiconductor &  0.02 $|$  0.04 $|$  0.55  & -0.64    \\ 
37 & $ \rm{Ta_2Se_2} $ & 4.03 & 10.20 & 8.65  & 7.51 & 11.35 & 2.70 & 2.59 & Metal &  ----- & -0.27  \\ 
38 & $ \rm{W_2Se_2} $ & 3.95 & 10.02 & 8.83  & 7.46 & 11.40 & 2.57 & 2.66 & Metal & -----  & 0.16    \\
39 & $ \rm{Re_2Se_2} $ & 3.93 & 9.97 & 8.88  & 7.51 & 11.35 & 2.46 & 2.65 & Semimetal &  ----- & 0.20   \\ 
40 & $ \rm{Ir_2Se_2} $ & 3.41 & 10.43 & 8.42  & 7.04 & 11.82 & 2.40 & 3.39 & Semiconductor & 0.03 $|$  0.08 $|$  0.16   & -0.04   \\  \midrule

\addlinespace

41 & $ \rm{Ti_2Te_2} $ & 4.27 & 10.27 & 8.59  & 7.43 & 11.42 & 2.84 & 2.72 & Magnetic-Metal & -----  & -0.43  \\ 
42 & $ \rm{V_2Te_2} $ & 3.97 & 10.11 & 8.75  & 7.40 & 11.46 & 2.71 & 2.66 & Metal & -----  & -0.12   \\ 
43 & $ \rm{Cr_2Te_2} $ & 3.79 & 9.95 & 8.91  & 7.37 & 11.48 & 2.58 & 2.67 & Magnetic-Metal & -----  & 0.12 \\ 
44 & $ \rm{Mn_2Te_2} $ & 3.74 & 9.94 & 8.92  & 7.43 & 11.43 & 2.51 & 2.62 & Magnetic-Metal & -----  & 0.07   \\ 
45 & $ \rm{Fe_2Te_2} $ & 3.62 & 9.98 & 8.87  & 7.39 & 11.47 & 2.59 & 2.56 & Magnetic-Metal & -----  & 0.06   \\ 
46 & $ \rm{Co_2Te_2} $ & 3.96 & 10.19 & 8.66  & 7.73 & 11.13 & 2.46 & 2.47 & Metal & -----  & -0.03    \\
47 & $ \rm{Ni_2Te_2} $ & 3.89 & 10.14 & 8.72  &7.62 & 11.24 & 2.50 & 2.52 & Metal & -----  & -0.17   \\    
48 & $ \rm{Cu_2Te_2} $ & 3.90 & 10.09 & 8.77  & 7.45 & 11.40 & 2.63 & 2.61 & Metal &  ----- & -0.05   \\ 
49 & $ \rm{Zn_2Te_2} $ & 4.33 & 10.36 & 8.49  & 7.55 & 11.31  & 2.67 & 2.81 & Semiconductor & 0.91 $|$  0.54 $|$  1.29   & -0.38    \\   
50 & $ \rm{Zr_2Te_2} $ & 4.54 & 10.36 & 8.50  & 7.37 & 11.48 & 2.98 & 2.85 & Semiconductor & 0.01 $|$  0.03 $|$  0.55  & -0.50    \\ 
51 & $ \rm{Nb_2Te_2} $ & 4.22 & 10.19 & 8.67  & 7.31 & 11.54 & 2.87 & 2.79 & Metal &  ----- & -0.14   \\ 
52 & $ \rm{Mo_2Te_2} $ & 4.06 & 10.01 & 8.84  & 7.29 & 11.56 & 2.72 & 2.81 & Metal & -----  & 0.11    \\ 
53 & $ \rm{Ru_2Te_2} $ & 4.13 & 10.21 & 8.65  & 7.62 & 11.24 & 2.59 & 2.60 & Metal & -----  & 0.15    \\ 
54 & $ \rm{Tc_2Te_2} $ & 4.04 & 9.99 & 8.86  & 7.40 & 11.45 & 2.59 & 2.75 & Metal & -----  & 0.11   \\  
55 & $ \rm{Cd_2Te_2} $ & 4.67 & 10.48 & 8.38  & 7.43 & 11.42 & 2.86 & 3.04 & Semiconductor & 1.02 $|$  0.63 $|$  1.33  & -0.37  \\ 
56 & $ \rm{Hf_2Te_2} $ & 4.38 & 10.33 & 8.53  & 7.27 & 11.59 & 3.06 & 2.82 &  Magnetic-Metal & -----  & -0.25  \\ 
57 & $ \rm{Ta_2Te_2} $ & 4.19 & 10.18 & 8.68  & 7.30 & 11.56 & 2.88 & 2.78 & Metal & -----  & 0.06   \\  
58 & $ \rm{W_2Te_2} $ & 4.06 & 10.00 & 8.85  & 7.25 & 11.61 & 2.76 & 2.84 & Metal & -----  & 0.35   \\ 
59 & $ \rm{Re_2Te_2} $ & 3.94 & 9.97 & 8.89  & 7.23 & 11.63 & 2.74 & 2.81 & Metal &  ----- & 0.34    \\ 
60 & $ \rm{Ir_2Te_2} $ & 4.16 & 10.28 & 8.58  & 7.64 & 11.22 & 2.58 & 2.64 & Metal & -----  & 0.08   \\ 

				\bottomrule
			\end{tabular}
			\begin{tablenotes}
				\small
				\vspace{1pt}
				\item[*]  $a=b$  \
				\hspace{20pt}
				 \item[$\dagger$] \ $\textnormal d^{\mathstrut}_{\textnormal{M1}\mbox{-}\textnormal{X1}}=\textnormal d^{\mathstrut}_{\textnormal{M2}\mbox{-}\textnormal{X2}}$ 
				\hspace{10pt}
				\item[$\ddagger$] \
$\textnormal d^{\mathstrut}_{\textnormal{M1}\mbox{-}\textnormal{X2}}=\textnormal d^{\mathstrut}_{\textnormal{M2}\mbox{-}\textnormal{X1}}$

			\end{tablenotes}
			
			\label{table:struct}
	\end{threeparttable}
\end{table*}

\section{Results and discussion}\label{sec3}

\subsection{Stability of M$_2$X$_2$ monolayers}

Let us start by discussing the energetics and stability of  M$_2$X$_2$ monolayers. Our calculations encompass a wide range of compounds, specifically M$_2$X$_2$ where M represents TM groups 3 to 11 elements and X represents S, Se, and Te. First, let us take a look at the formation energy $E_\textnormal f$ and the factors that influence it for M$_2$S$_2$ compounds. The unit cell of M$_2$S$_2$ consists of two M and two X atoms. The formation energies are determined using the equation $E_\textnormal f$$=$$E_{\textnormal{tot}}$(M$_2$X$_2$)$-$$2E$(M)$-$$2E$(X), where $E_{\textnormal{tot}}$(M$_2$X$_2$) represents the total energy of 2D M$_2$X$_2$ per unit cell, and $E$(M) and $E$(X) denote the total energy of M and X per atom, respectively.
The total energies of the M elements are derived from their most stable bulk structures.
For X atoms (e.g., S, Se, Te), $E$(X) is determined based on the total energies of their native elemental forms. 
For example, for sulfur it corresponds to orthorhombic $\alpha$-sulfur which is the native and most stable form of sulfur under standard conditions.
This approach allows for the calculation of formation energies by considering the total energies of the constituent atoms and the compound itself within the unit cell.

 There are two dominant competing factors to the understanding of the formation energies of this class of
compounds. The first factor involves the expansion of the TM lattice due to the insertion of chalcogen X atoms, and the second factor pertains to the hybridization between the valence $\textnormal d$ states on the TM and the $\textnormal p$ states of the chalcogen X.
When the chalcogen X atoms are inserted into the TM lattice, the $\textnormal d$-band is narrowed, leading to a reduction in the $\textnormal d$-band broadening contribution to the formation energy. The decrease in $\textnormal d$-bond energy is closely related to the volume of the chalcogen X and the strength of the $\textnormal d$ bonding in the TM constituent.
In Table\,\ref{table:struct}, we report the calculated formation energies $E_\textnormal f$ for the M$_2$S$_2$, M$_2$Se$_2$, and M$_2$Te$_2$ phases. In all compounds, the formation energies obey the relation $E_\textnormal f$(M$_2$S$_2$)$<$$E_\textnormal f$(M$_2$Se$_2$)$<$$E_\textnormal f$(M$_2$Te$_2$), suggesting that the M$_2$S$_2$ phases are more likely to be achieved with greater purity in experimental settings.
The atomic radius of chalcogen increases as one move from X=S to Te. Consequently, the necessary volume expansion energy to accommodate S atoms will be lower regardless of the type of TM atoms. In other words, if the TM atoms are pulled farther apart, the $\textnormal d$-bandwidth decreases, leading to a reduction in the $\textnormal d$-band broadening contribution to the stability of the lattice. This is the main reason for the lower
formation energies of the M$_2$S$_2$ compounds. 

Additionally, as one progresses across the row from Ti to Cu, interatomic distance in original TM structures initially reduces up to the Cr atom beyond which it remains nearly constant up to the Co atom, and finally slightly increases to Zn.  So, the required volume expansion energy to accommodate any of the
X elements exhibits a similar behavior. This is the reason that among the considered M$_2$S$_2$, Ti possesses the smallest formation energy. 
3$\textnormal d$ TM atoms have a smaller interatomic distance than 4$\textnormal d$ and 5$\textnormal d$ TM atoms. As a result, hosting X will give a weaker expansion effect in 3$\textnormal d$-based M$_2$S$_2$ systems.  This more or less agrees with our calculated formation energy when we move from 3$\textnormal d$ to 5$\textnormal d$ compounds. The exceptions such as Zr$_2$S$_2$ are related to the effect of $\textnormal p-\textnormal d$ hybridization which acts as a competing effect.
When a compound is formed from a TM atom and a X element with a valence p shell, the $\textnormal d$ states of the TM hybridize with the $\textnormal p$ states of X to create bonding and antibonding hybrid states.
This hybridization leads to a effective term in the heat of formation for a few compounds.

By analyzing the phonon dispersion and checking for negative frequencies, we can assess the stability of materials and identify any potential dynamical instabilities.
Twenty-seven compounds are expected to be stable based on the phonon calculations (See Fig. S1, Fig. S2, and Fig. S3  of  supplementary file) whose electronic structures will be discussed later.
The other compounds may be dynamically unstable since they exhibit imaginary phonon branches.
Here, we have presented the phonon spectra for selected materials, that is M$_2$S$_2$ (M=Zr, Re, Tc) and Mn$_2$Se$_2$ in Fig.\,\ref{fig-phonon}. The results reveal that almost all phonon frequencies are positive, indicating that these four systems are in their equilibrium state. Consequently, it is anticipated that under suitable conditions various M$_2$X$_2$ monolayers maybe realized. 

Additionally, studying the thermal stability of a material is crucial for understanding energy fluctuations over time at various temperatures. The evolution of energy during the simulation and the final snapshots of the resulting geometries at 300 K of Tc$_2$S$_2$ and Zn$_2$S$_2$, illustrated in Figs.\,\ref{fig-AIMD}(a) and \,\ref{fig-AIMD}(b) respectively, is assessed through AIMD calculations. 
Both materials show no significant changes in their energy spectra, indicating that there are no broken bonds or geometric structure reconstructions after heating the system for 6500 fs. 
Time-dependent temperature fluctuation of Tc$_2$S$_2$ and Zn$_2$S$_2$ by AIMD simulations are also shown in Fig.\,\ref{fig-AIMD}.
This observation confirms the thermodynamic stability of the M$_2$X$_2$ systems.

\begin{figure*}[!h]
\begin{center}
\includegraphics[scale=0.38]{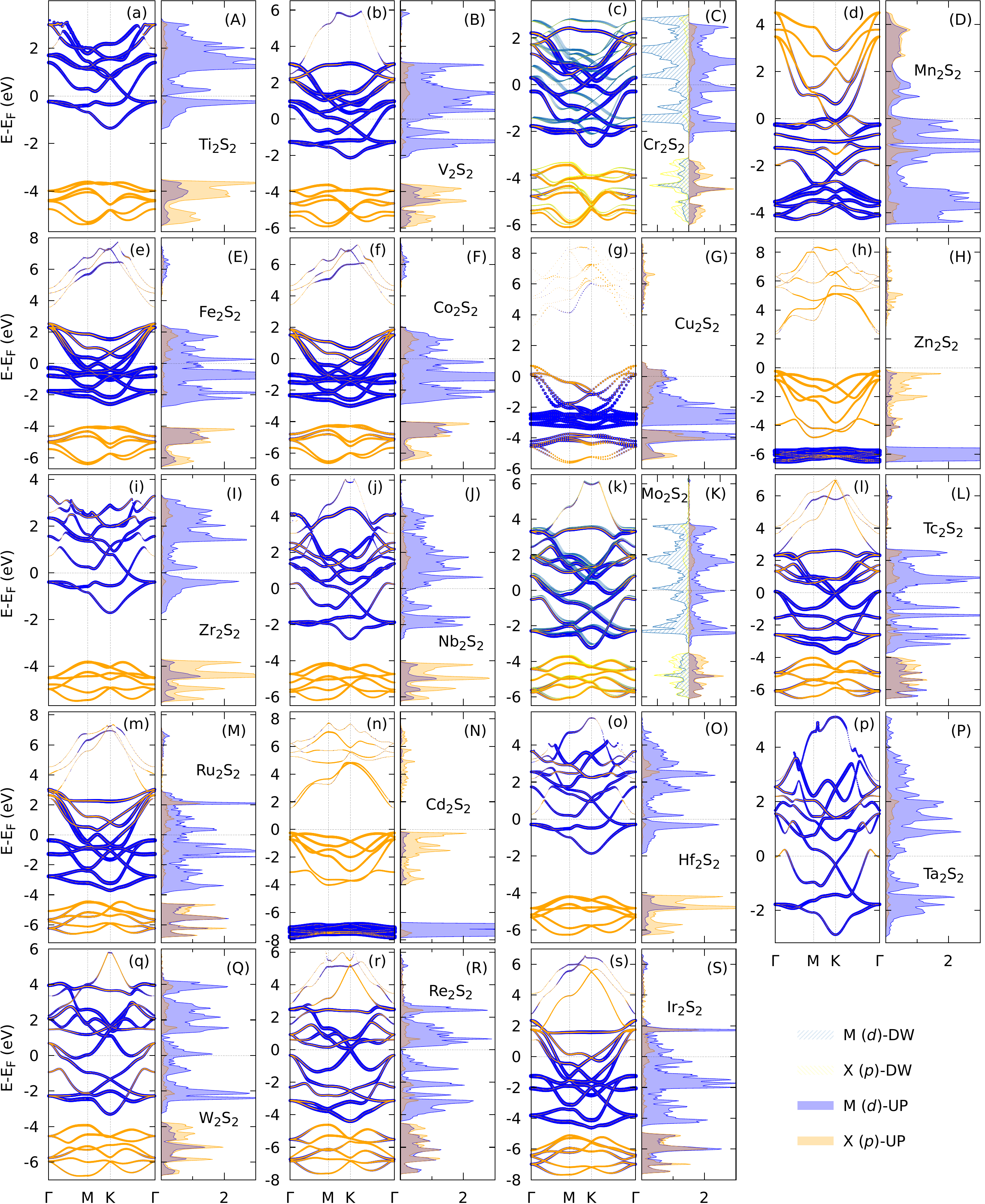}
\end{center}
\vspace*{-0.5cm} 
\caption{(Colors online) Band structure and DOS projected onto $\textnormal d$ states of the M atom (blue color) as well as on $\textnormal p$ states of the X atoms (orange color) for M$_2$S$_2$ materials. For spin-polarized calculations, dark color (light color) correspond to spin-up (spin-down) states.} 
\label{fig-bands}
\end{figure*}

\subsection{Mechanical properties}

Now, we explore the mechanical characteristics of stable structures by examining elastic coefficients, in-plane stiffness \(Y_{\textnormal{2D}}\), shear modulus $G$, and Poisson’s ratio \( \nu \). By analyzing the elastic strain tensors \(C_{ij}\), we can derive the mechanical properties and assess elastic stability. The calculated values of \(C_{ij}\) are reported in Table \ref{table:mechanics}. For the 2D hexagonal systems we designed, elastic stability is confirmed when all \(C_{ij}\) values are positive and obey the Born and Huang criteria~\cite{Born,Mouhat}, which stipulate that $C_{11}$ $>$ $|C_{12}|$.
Table \ref{table:mechanics} indicates that structures exhibiting dynamic stability also demonstrate mechanical stability. In the context of two dimensions, \(Y_{\textnormal{2D}}\) quantifies the rigidity or flexibility of a crystal when subjected to external loads, calculated using the formula: 
$Y_{\textnormal {2D}}$ = ($C_{11}^{2}$ $-$ $C_{12}^{2}$)/$C_{11}$. The calculated $Y_{\textnormal{2D}}$ values for the monolayers of Zr\(_2\)S\(_2\), V\(_2\)Se\(_2\), Re\(_2\)S\(_2\), and Tc\(_2\)S\(_2\) are 89.1, 72.2, 143.5, and 91.2 N m\(^{-1}\), respectively.   Due to the symmetric and isotropic nature of the crystal structures analyzed, the \(Y_{\textnormal{2D}}\) values in the \(x\) direction are equivalent to those in the \(y\) direction.

\begin{table}[!ht]
	\centering
	\begin{threeparttable}
		\caption{Relaxed-ion elastic coefficients $C_{ij}$, 2D Young’s modulus $Y_{\textnormal {2D}}$, shear modulus $G$,  and Poisson’s ratio \( \nu \) for stable M$_2$X$_2$ structures.} 
			\begin{tabular}{cccccccc}
				\toprule
				 M$_2$X$_2$ & $C_{11}=C_{22}$ & $C_{12}$ & $Y_{\textnormal {2D}}$ & $G$=($C_{11}$-$C_{12}$)/2 & \( \nu \)  \\
                                 & (N m$^{-1})$  & (N m$^{-1})$ & (N m$^{-1})$ & (N m$^{-1})$ & ($-$)   \\
\\[-2.1ex]
\midrule  \addlinespace

$ \rm{Ti_2S_2} $ & 110.9 &  49.6 & 88.8  &  30.7  & 0.45  \\  
$ \rm{Ti_2Se_2} $ &  90.8 & 42.1 & 71.3  &  24.4  & 0.46  \\  
$ \rm{V_2Se_2} $ & 100.8  &  53.7 & 72.2  &  23.6  & 0.53 \\  
$ \rm{V_2Te_2} $ & 83.3  &  44.9 &  59.1  &  19.2  & 0.54 \\  
$ \rm{Mn_2S_2} $ & 98.0 &  18.0 & 94.7  &  40.0  & 0.18 \\
$ \rm{Mn_2Se_2} $ & 90.7 &  19.3 & 86.6  &  35.7  & 0.21 \\
$ \rm{Fe_2S_2} $ & 148.6  &  53.0 & 129.7  &  47.8  & 0.36 \\ 
$ \rm{Fe_2Se_2} $ & 102.9  &  27.6 &  95.5  &  37.6  & 0.27 \\  
$ \rm{Fe_2Te_2} $ & 105.1  &  53.8 &  77.6  &  25.7  & 0.51 \\  
$ \rm{Co_2S_2} $ & 138.4  &  52.8 & 118.3  &  42.8  & 0.38\\ 
$ \rm{Co_2Se_2} $ & 112.9  &  43.1 & 96.5  &   34.9  & 0.38\\ 
$ \rm{Co_2Te_2} $ & 98.0  &   33.0 & 86.9  &   32.5  & 0.34\\ 
$ \rm{Cu_2S_2} $ & 92.3  &  35.8 & 78.4  &  28.2  & 0.39\\   
$ \rm{Cu_2Se_2} $ & 68.6  &  29.5 & 55.9  &  19.5 & 0.43\\
$ \rm{Cu_2Te_2} $ & 66.8  &  29.3 & 54.0  &  18.8 & 0.44\\
$ \rm{Zn_2S_2} $ & 98.2  &  27.7 &  90.4 &  35.2  & 0.28 \\ 
$ \rm{Zn_2Se_2} $ & 79.1  &  23.2 &  72.3  &  27.9  & 0.29 \\  
$ \rm{Zn_2Te_2} $ &  66.4 &  19.2 &  60.8   &  23.6  & 0.29 \\  
$ \rm{Zr_2S_2} $ &  114.3  &  53.8  &  89.1   &  30.3  & 0.47 \\ 
$ \rm{Zr_2Se_2} $ &  95.8 &  46.6 &   73.1   &  24.6  & 0.49 \\ 
$ \rm{Nb_2Te_2} $ &  88.2 &  52.2 &   57.3   &  18.0  & 0.59 \\ 
$ \rm{Tc_2S_2} $ & 126.5  &  66.9 & 91.2 & 29.8  & 0.53     \\
$ \rm{Tc_2Se_2} $ & 132.2 &  51.3 & 112.3  & 40.4 & 0.39     \\ 
$ \rm{Tc_2Te_2} $ & 102.6 &   53.0 & 75.2  &  24.8 & 0.52   \\ 
$ \rm{Ru_2S_2} $ &  146.9  & 60.5 & 122.1  & 43.3 & 0.41    \\ 
$ \rm{Ru_2Se_2} $ & 119.9  & 54.6 & 94.9 & 32.6 & 0.46    \\ 
$ \rm{Ru_2Te_2} $ & 102.1  & 50.5 & 77.1 & 25.8 & 0.49   \\ 
$ \rm{Cd_2S_2} $ & 79.6  & 25.1 & 71.7 & 27.2 & 0.32    \\ 
$ \rm{Cd_2Se_2} $ & 65.9  & 21.7 & 58.8 &  22.1 & 0.33   \\ 
$ \rm{Cd_2Te_2} $ & 54.3  & 17.3 & 48.9  &  18.5 & 0.32   \\ 
$ \rm{Hf_2S_2} $ & 127.4 &  61.8 & 97.5 & 32.8 & 0.49     \\  
$ \rm{Ta_2Te_2} $ & 99.4 &   60.7 & 62.4 & 19.4 & 0.61     \\ 
$ \rm{Re_2S_2} $ & 169.5 & 66.4 & 143.5  & 51.6  & 0.39     \\ 
$ \rm{Re_2Se_2} $ & 148.3 &  59.9 & 124.1  &  44.2  & 0.40  \\  
$ \rm{Ir_2Se_2} $ & 127.3 &  57.6 & 101.5 & 34.9  & 0.45     \\
\bottomrule
\addlinespace
			\end{tabular}			
			\label{table:mechanics}
	\end{threeparttable}
\end{table}

Comparison of \(Y_{\textnormal{2D}}\) with other well known 2D materials, namely graphene (341.1 N m$^{-1}$), MoSi$_2$N$_4$ (491.5 N m$^{-1}$), MoS$_2$ (124.5 N m$^{-1}$)  and  $h$-BN (275.9 N m$^{-1}$), reveals that M$_2$X$_2$ materials are mechanically as stable 
as MoS$_2$ but less stiffer than MoSi$_2$N$_4$  and single layer $h$-BN~\cite{Sahin,Cakir,Bafekry}.
To elucidate the impact of applied stress on the mechanical properties of the systems, we determined the Poisson's ratio \( \nu \) of the materials. The Poisson's ratio is defined as the negative ratio of transverse contraction strain to longitudinal extension strain in the direction of the stretching force. Utilizing the elastic tensor \( C_{ij} \), \( \nu \) is calculated using the formula $\nu =C_{11}/C_{12}$. The obtained results fall within the range of 0.29 to 0.54, as presented in Table \ref{table:mechanics}. For comparison, the reported values of \( \nu \) for MoSi\(_2\)N\(_4\) and MoS$_2$ crystals are 0.28 and 0.25, respectively~\cite{Cakir,Bafekry}.

\subsection{Electronic and magnetic properties}

In this section, we present the resultsof our investigation of the electronic and magnetic properties of  
each M$_2$S$_2$ systems. Electronic structures reveal that M$_2$S$_2$ can be classified into three distinct categories 
depending on TM atoms. These categories are referred to as metals, semiconductors/semimetals, and magnetic metals. 
The projected band structures depicted in Fig.\,\ref{fig-bands} (see also Figs. S4 to S6 in supplementary material) show that compared to the other orbitals from chalcogen atoms, the $\textnormal d$ orbitals of
the TM atom significantly contributes to the bands near the Fermi level (with some exceptions). 
Thus, similar to other stable TM-based materials such as TMDs and MXenes~\cite{Ramezani,Yekta}, most of the investigated M$_2$S$_2$ compounds can be described by an effective low-energy model based on only  $\textnormal d$  electrons of TM atom.  Additionally, the $\textnormal p$-bands of X atoms appear below the $\textnormal d$-bands of 
the TMs separated by a large energy difference.

\subsubsection{Non-magnetic metals}

Our first-principles calculations indicate that the most stable M$_2$X$_2$ systems are metallic. Fig.\,\ref{fig-bands} shows projected band structure and densities of states
(PDOS) of M$_2$S$_2$ compounds. Additionally, we reported electronic structure of all other M$_2$X$_2$ compounds in supplementary file. 
For instance, in the case of V$_2$S$_2$, the states around the $E_\textnormal F$ of these compounds are primarily dominated by the $\textnormal d$ orbitals of the V atoms, as shown in Fig.\,\ref{fig-bands}(b). The compounds exhibit common patterns of hybridization, with bonding states between the $\textnormal d$ orbitals of M and the $\textnormal p$ orbitals of X located below $E_\textnormal F$. 
Above these bonding states, non-bonding states of M are located near the $E_\textnormal F$. 
In V$_2$S$_2$, the p states of X atoms are partially hybridized with M $\textnormal d$-orbitals below the $E_\textnormal F$ between -4.0 and -6 eV, and they are separated by a 2 eV band gap from the $\textnormal d$-bands of V.
From Fig.\,\ref{fig-bands}(a) to Fig.\,\ref{fig-bands}(g), admixture of chalcogen $\textnormal p$ with $\textnormal d$ states increase as
one moves from M=Ti to Cu-based systems.
Indeed, as we move from Ti to Cu, the chalcogen $\textnormal p$ states gradually move to the higher energy and are accompanied by TM $\textnormal d$ bands. 
In Fig.\,\ref{fig-bands}, moving again from left Zr to right Ru for 4$\textnormal d$ TM in the periodic table, 
(Hf to Tc for 5$\textnormal d$ TM), the same increasing trend is observed for X$-$$\textnormal p$ and M$-$$\textnormal d$ energy separation. 
S, Se, and Te have same valence electrons in their last electronic shells. Hence, the M$_2$Se$_2$, and M$_2$Te$_2$ 
compounds follow almost the same electronic structure as the M$_2$S$_2$ systems.
Ongoing from S to Te within each M$_2$X$_2$ system, the lattice
constant increases (see Table\,\ref{table:struct}), as a consequence, the longer bond
lengths lead to larger dispersion for bands with X$-$$\textnormal p$ character.  It can bring the
states energetically closer together as shown in Fig. S6 of supplementary material for
the Te-based structure.
Such smaller energy difference in the M$_2$S$_2$-M$_2$Se$_2$-M$_2$Te$_2$ sequence has been observed 
in some other 2D TM materials such as TMDs and TM-halides~\cite{Yasin,Karbala}. 

\subsubsection{Semimetals and semiconductors}
While most of the M$_2$X$_2$ have metallic behavior, a few of them such as 
M$_2$X$_2$(M=Zn, Cd) are direct band-gap semiconductors (Fig.\,\ref{fig-bands}(h) and Fig.\,\ref{fig-bands}(n)). 
Since the simple GGA method is known to potentially underestimate band gap values, the computationally more expensive HSE functionals are utilized to enhance the accuracy of band gap estimation.
We report the sizes of the band gaps of all the semiconducting monolayers in
Table \,\ref{table:struct} based on both the standard DFT+PBE and HSE06 calculations.
Some applications of these compounds include electronic devices, photonic and energy harvesting devices, 
novel spin qubit frameworks for quantum computation applications, and solar cells. 
The computed energy gap $E_\textnormal g$ varies in the range 0.9–2.6 eV (1.3–3.7 eV) at the DFT+PBE (HSE06) level. 
As seen in Table \,\ref{table:struct}, the $E_\textnormal g$ tends to reduce when X is varied from S to Te.

Fig.\,\ref{fig-bands}(h) displays the corresponding projected electronic structure for
the Zn$_2$S$_2$, which highlights the significant contribution of the 4$\textnormal p$ orbital of Zn and 3$\textnormal p$ orbital of S atoms to the
valence band region near the $E_\textnormal F$, while
the 5$\textnormal s$ orbital of Zn dominates in the conduction band region.
The semiconducting nature can be understood based on the structural properties and the electron filling of outer shell.
Here, after bond formation, the 3$\textnormal d$ orbitals of Zn atoms are expected to be nearly full due to 
charge compensation of each atom which makes this system semiconducting.

\begin{figure}[!h]
\begin{center}
\includegraphics[scale=0.48]{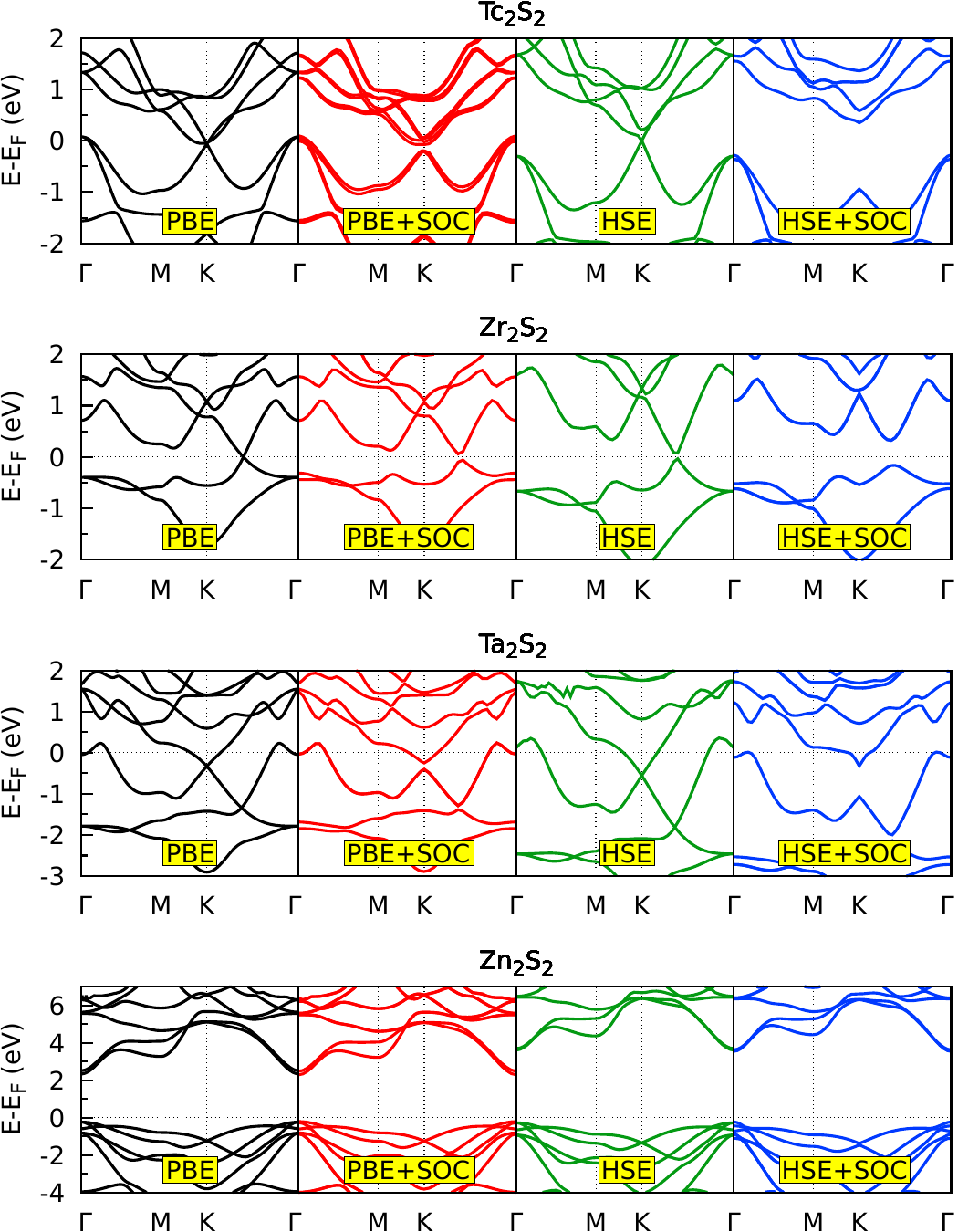}
\end{center}
\vspace*{-0.5cm} 
\caption{(Colors online) PBE, PBE+SOC, HSE06, and SOC+HSE06 band structure of (a) Tc$_2$S$_2$, (b) Zr$_2$S$_2$, (c) Ta$_2$S$_2$, and (d) Zn$_2$S$_2$. The Fermi level is set to zero energy.} 
\label{SOC+HSE}
\end{figure}

Beside these wide band-gap semiconductors, M$_2$X$_2$(M=Ti, Zr, Hf) and M$_2$X$_2$(M=Tc, Re) become semimetals in this family with more or less similar energy dispersion based on GGA+PBE method.
This similar behavior among Ti, Zr, and Hf (same as Tc and Re) can be attributed to the fact that they belong to the same group in the periodic table and have the same number of electrons in their outermost atomic orbital shells. As a result, M$_2$X$_2$ systems derived from these elements demonstrate similar semimetallic behavior when placed in the same crystal structure.
This semimetallic behavior (cone-like shape)  \cite{Cone,borophene} has not been observed in other 2D materials such as TMDs, MXenes, TM halides, and intercalated architecture MA$_2$Z$_4$ and therefore remains an exclusive property of this family. 
Note that the top of valence-band of M$_2$X$_2$ (M=Ti, Zr, Hf) compounds touch conduction-band at k-point along 
K$\Gamma$ path, while in the case of M$_2$X$_2$(M=Tc, Re)  the band crossing happens at high symmetry $K$ point.   
The electronic band structures of M$_2$S$_2$ (M=Zr, Tc, Ta, Zn),  are calculated with consideration of the SOC, as depicted in Fig.\,\ref{SOC+HSE}. 
In all cases, a strikingly similar behavior is observed in the band structures calculated with and without SOC contributions. In the case of Zr$_2$S$_2$ and Tc$_2$S$_2$, it is noted that the conduction-band minimum (CBM) slightly shifts upwards, while the valence-band maximum (VBM) shifts downwards, resulting in a narrow band gap opening of approximately 0.1 eV due to the presence of SOC making them potentially useful as infrared detectors/sensors. 
Our HSE calculations indicate that M$_2$X$_2$ (M=Ti, Zr; X=S, Se) are a direct band-gap semiconductor with 
$E_\textnormal g$ between 0.49 to 0.99 eV.  The band dispersion and band-gap opening  for compounds
with cone-like bands at high symmetry $K$ point (in the level of GGA+PBE) is different and more complicated.
For example in the case of Tc$_2$S$_2$, system is still semimetal in HSE calculation. We found that in both PBE+SOC and HSE+SOC calculation, the band gap is indirect and is significantly larger than corresponding one
for Zr$_2$S$_2$.

\begin{figure}[!h]
\begin{center}
\includegraphics[scale=0.41]{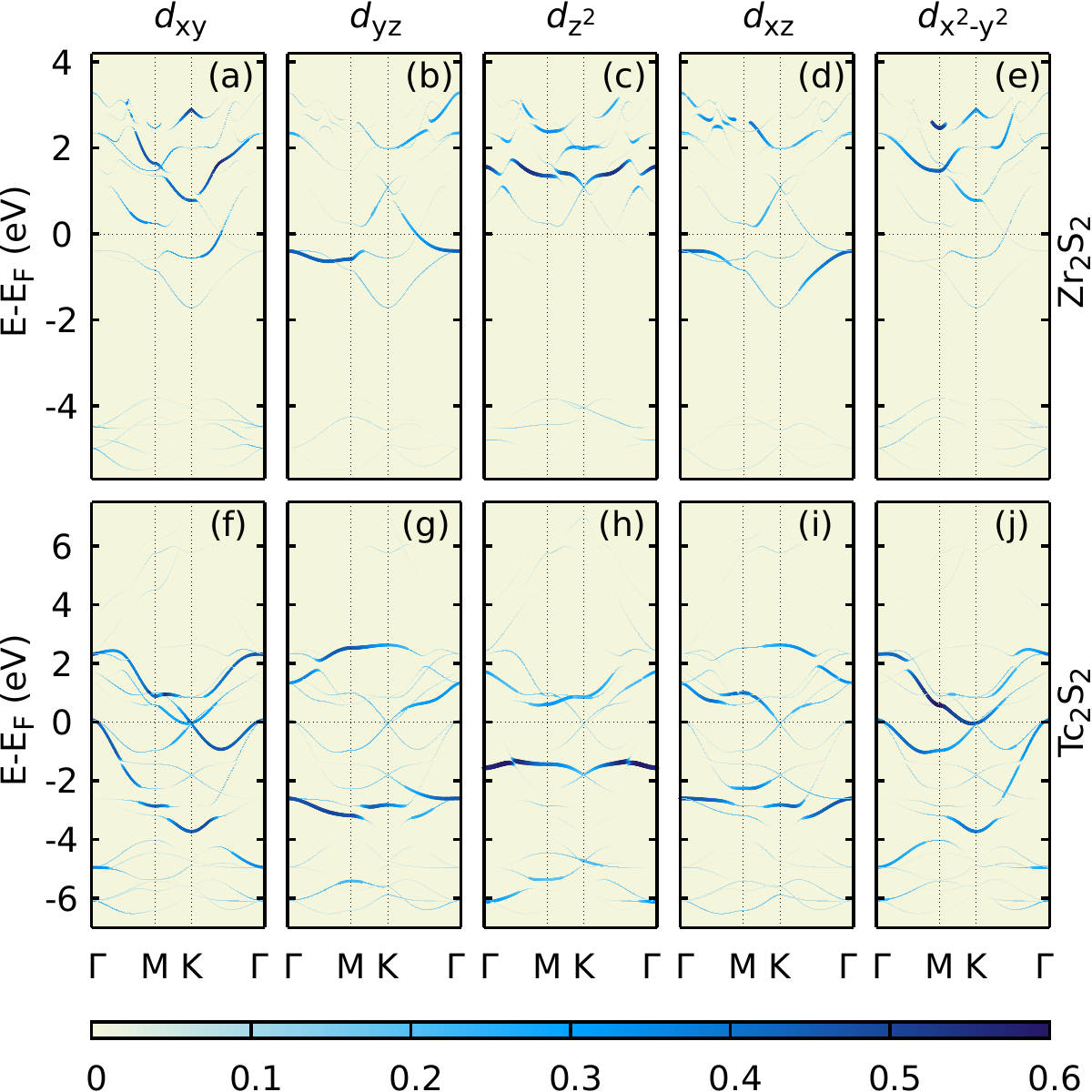}
\end{center}
\vspace*{-0.5cm} 
\caption{(Colors online) The orbital-projected band
structure for $\textnormal d$ electrons of TM atom of (a)-(e) Zr$_2$S$_2$ and (f)-(j) Tc$_2$S$_2$ based on DFT-PBE.} 
\label{d-orbitals}
\end{figure}

To further analyze the orbital character in band structure of semimetallic systems, we have investigated the projected band structure of the Zr$_2$S$_2$ and Tc$_2$S$_2$, as shown in Figs.\,\ref{d-orbitals}(a)-\,\ref{d-orbitals}(e) and Figs.\,\ref{d-orbitals}(f)-\,\ref{d-orbitals}(j) respectively. One can easily find that VBM and CBM for the Zr$_2$S$_2$ are formed by the $\textnormal d_{xz}$/$\textnormal d_{yz}$ and $\textnormal d_{xy}$/$\textnormal d_{x^2-y^2}$/$\textnormal d_{z^2}$ states of TM atoms respectively. This confirms the symmetry discussion that $d_{xz}$ and $d_{yz}$ are farther from the ligands than the others and therefore experiences 
less repulsion. Note that the valence bands (conduction bands) are not of  pure $\textnormal d_{xz}$/$\textnormal d_{yz}$ ($\textnormal d_{xy}$/$\textnormal d_{x^2-y^2}$/$\textnormal d_{z^2}$) 
character and the symmetry allows mixtures from from X$-$$\textnormal p$ states. This denominations thus refer
to their dominant orbital character.  In the case of Tc$_2$S$_2$, the $\textnormal d_{xz}$/$\textnormal d_{yz}$ states are almost occupied and the bands 
with $\textnormal d_{xy}$/$\textnormal d_{x^2-y^2}$ character are dominant at the vicinity of $E_\textnormal F$ that participate in the formation of the Dirac cone.

\begin{figure}[!h]
\begin{center}
\includegraphics[scale=0.35]{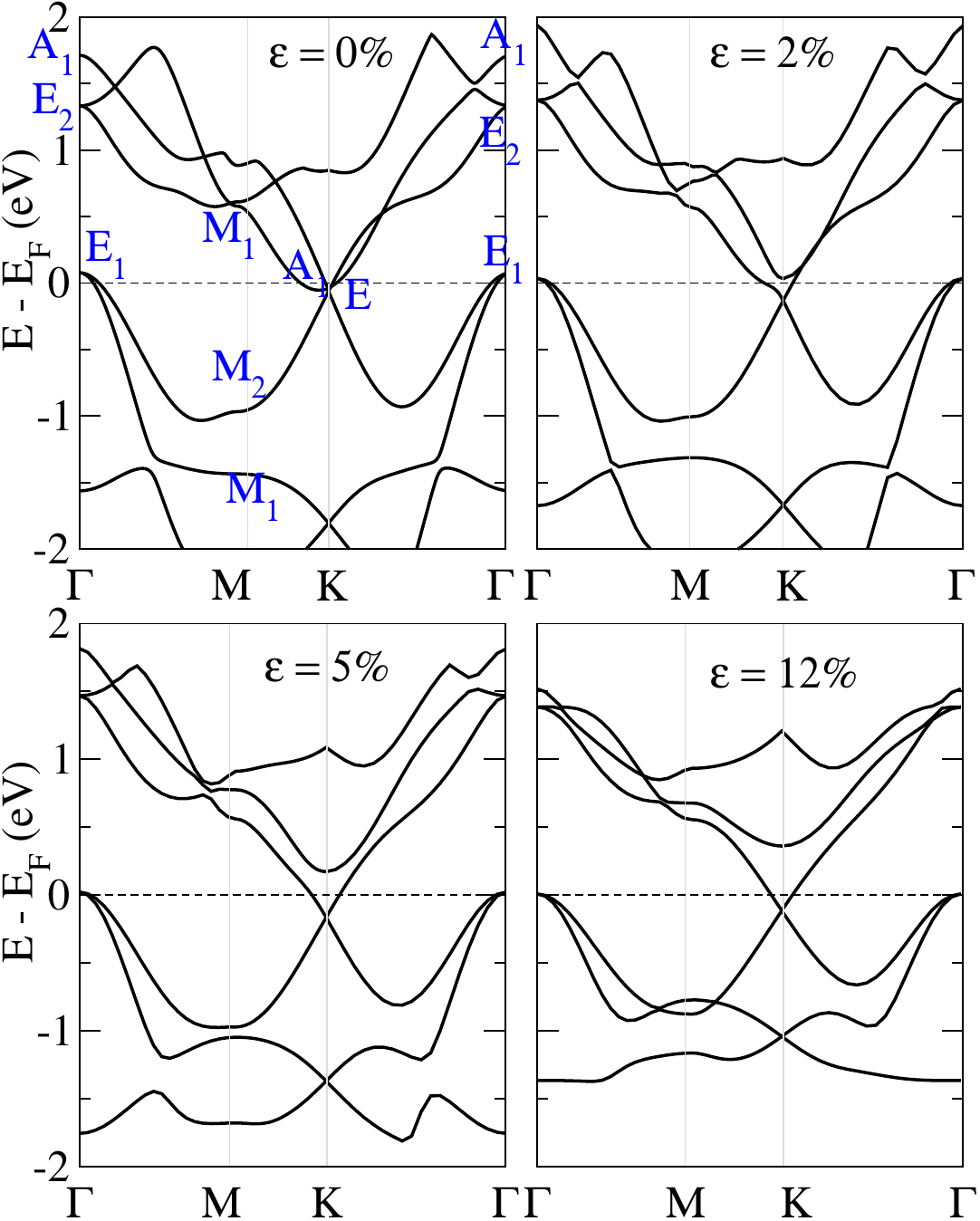}
\end{center}
\vspace*{-0.5cm} 
\caption{(Colors online) The total band
structure for Tc$_2$S$_2$ strained structures with (a) free state (not strain) (b) 2$\%$ tensile strain, (c) 5$\%$ tensile strain, and (d) 12$\%$ tensile strain based on DFT-PBE.} 
\label{fig:press}
\end{figure}

Since the Dirac cone in Tc$_2$X$_2$ and Re$_2$X$_2$ occurs at the $K$ point similarly to the well known graphene, where relativisitc fermions aspects are combined with the strongly correlated aspects in the same system. Therefore they hold a great promise as a platform for correlated relativistic fermions allowing to explore combination of topological and many-body aspects. Let us list the two conceptual difference between Tc$_2$S$_2$ and graphene. Firstly, the $\textnormal d$-states participate in the formation of the Dirac cone, whereas in graphene, the carbon $\textnormal p_z$-states play this role. Secondly  the two bands meeting at Fermi level and forming the Dirac cone in Tc$_2$S$_2$ are not isolated from other bands.
As shown in Fig.\,\ref{d-orbitals}(f) and Fig.\,\ref{d-orbitals}(j), a quadratic band at $E_\textnormal F$ with $\textnormal d_{xy}$/$\textnormal d_{x^2-y^2}$ character coexists with two $\textnormal d_{xz}$/$\textnormal d_{yz}$ bands.
Therefore, we expect that applying biaxial pressure will couple with the $\textnormal d_{xy}$/$\textnormal d_{x^2-y^2}$ states and create significant band structure variation. This can serve as a convenient tool for the engineering of such a rich Dirac-Schr\"odinger bands. 
For these reasons, we examined the effect of strain on the electronic properties of Tc-based system. 
Fig.\,\ref{fig:press} displays the band structure of Tc$_2$S$_2$ for different tensile strain.  
As seen, the quadratic band at the Fermi surface moves away from the Dirac bands with a moderate strain of about 5$\%$, resulting in a clearly defined Dirac cone in the Tc$_2$S$_2$. This is a unique platform that allows to study the interplay between Dirac and quadratic bands where the coupling between the relativistic fermions in the Dirac bands and non-relativistic fermions in the quadratic bands can be tuned by the external influence of the strain. 

\subsubsection{Magnetic metals}

The number of  2D materials that naturally display magnetic ordering in their pristine form is quite limited. Among the
vast number of 2D materials, only MX$_3$(M=V, Cr; X=Br, I)~\cite{Seyler-CrI3,Siena,Kong}, Cr$_2$Ge$_2$Te$_6$~\cite{Gong-Cr2Ge2Te6}, and MX$_2$
(M=V, Mn;X=Se, Te)~\cite{Duvjir,Wang,Lasek,Purbawati,Wan,Meng} have been experimentally observed to exhibit magnetic orderings. 
Theoretical predictions have suggested the presence of intrinsic ferromagnetism in monolayers of M$_2$C MXenes like Ti$_2$X(X=C, N), Cr$_2$C, and V$_2$C~\cite{Amrillah,Yekta,Yue-Fe2C,Tan-V2C,Akgenc,Khazaei}, but this has not been experimentally confirmed.
Even for most of the 3$\textnormal d$-TMDCs like CrX$_2$ and MnX$_2$ systems, as well as VX$_2$,  the origin of the intrinsic
ferromagnetism remains controversial and the existence of a
magnetic phase is still debated~\cite{Duvjir,Feng,Chen-2018,Fumega,Wong}. In other words, conflicting studies have suggested that the observed ferromagnetism in 3$\textnormal d$-TMDCs is actually \emph{extrinsic} 2D magnetism arising from vacancies~\cite{Chua} or proximity effects~\cite{Vinai,Zhang-2019}, which cannot be definitively ruled out in the growth of 2D crystals. The emergence of such extrinsic magnetic moments and long-range magnetic order induced by atomic vacancies or edge states in other 2D nonmagnetic materials like graphene~\cite{Yazyev,Hanif,Ersoy,Ugeda,Bagherpour-2,Montaghemi} and MoS$_2$ ~\cite{Zhang-2013,Cai} makes it more plausible that pristine TMDCs lack intrinsic 2D magnetism.
The scarcity of magnetic materials can be attributed to the robust covalent bonding between the TM and the X element.
However, external strain can modify the covalent bonds, causing the release of $\textnormal d$ electrons and thereby giving rise to magnetism ~\cite{Yun-2012,Zhou-2012}.
For instance, when tensile strain increases in M$_2$C MXenes, the magnetic moments are significantly boosted, leading to a transition from nonmagnetic states to ferromagnetic state in parent nonmagnetic MXenes~\cite{Zhao-2014,Warner}. The most notable transition is observed in Hf$_2$C, where the magnetic moment rises to 1.5 $\mu_B$/unit at a strain of 1.8 $\%$~\cite{Zhao-2014}. Therefore finding new families of 2D materials possessing intrinsic magnetically non-trivial state is highly desirable.

In our M$_2$X$_2$ system, due to the the weak covalent bonding between  the TM and the X element, 
we expect more of these materials to be magnetic in M$_2$X$_2$ family compared to other TM-based materials. 
To identify the preferable magnetic ordering in M$_2$X$_2$ systems, we set the spin ordering to either ferromagnetic (FM) or antiferromagnetic (AFM), as shown in Fig. S7. 
The exchange energy ($E_{\textnormal{ex}}$) is calculated as $E_{\textnormal{ex}}=E_{\textnormal{AFM}}-E_{\textnormal{FM}}$. Positive exchange
energy indicates that the ground state of the system is ferromagnetic.
Our calculations suggest that the ground states of Cr$_2$S$_2$, M$_2$Te$_2$(M=Fe, Hf), and Ti$_2$Te$_2$ systems
are ferromagnetic, while Mn$_2$X$_2$(X=S, Se, Te),  Cr$_2$X$_2$(X=Se, Te),  Fe$_2$Se$_2$, and V$_2$Te$_2$ are antiferromagnetic as shown in Table\,\ref{table:magnetic}.  The non-positive values of the first column that gives the energy difference $E_{\rm m}=E_{\rm FM}-E_{\rm NM}$ indicates that non-magnetic states is now the lowest energy state. Then the second column decides whether the FM or AFM is the preferred state. As expected, the magnetic characteristics of these systems arise from the $\textnormal d$ orbitals of TM atoms. The magnetic moments of TM atoms in these systems are reported in Table\,\ref{table:magnetic}.

The value (also sign) of direct exchange and M–X–M superexchange can be controlled by the M–X–M bond 
angles as well as $\textnormal d$ electron configurations of TM atoms:
According to the  Goodenough–Kanamori–Anderson (GKA) formalism, in systems with 90$^{\circ}$ 
 bond angles the $\textnormal d$-orbitals on neighboring TM atoms overlap with different halogen-$\textnormal p$ orbitals and 
the superexchange interaction between TM atoms is always FM~\cite{Goodenough,Kanamori,Kulish}. 
In M$_2$X$_2$ compounds, all calculated metal-halogen-metal bond angles are between 60-65 degree.
In such a case, $\textnormal d$-orbitals on neighboring TM atoms are not able to  completely overlap with two 
orthogonal chalcogen-$\textnormal p$ orbitals.  So, FM superexchange is weakened due to deviation of the M–X–M bond angles from 90$^{\circ}$. 
On the other hand, the presence of 60$^{\circ}$  M–X–M bond angles causes the TM atoms to come closer to each other
and enhance the AFM direct exchange interaction.  That is why most of the magnetic M$_2$X$_2$ materials are 
in the AFM phase.

\begin{table}[!t]
	\centering
	\begin{threeparttable}
		\caption{The energy difference between
the total energy of ferromagnetic and non-magnetic phases $E_{\textnormal m}=E_{\textnormal{FM}}-E_{\textnormal{NM}}$, energy difference between
the total energy of antiferromagnetic and ferromagnetic configurations $E_{\textnormal{ex}}=E_{\textnormal{AFM}}-E_{\textnormal{FM}}$, the magnetic moment of TM atoms $\mu^{\textnormal{PBE}}$ in the unit of $\mu_\textnormal B$, Density of state at Fermi level $D(E_\textnormal F)$, effective Stoner parameter $I= (U + 6J)/5$, and exchange constant used in Heisenberg Hamiltonian $J_\textnormal H$.} 
			\begin{tabular}{ccccccccc}
				\toprule
				 M$_2$X$_2$ & $E_{\textnormal m}$ & $E_{\textnormal{ex}}$ & $\mu^{\textnormal{PBE}}$ & $D(E_\textnormal F)$ & $I$  & $J_\textnormal H$  \\
                                 & (meV)  & (meV) & ($\mu_\textnormal B$) & (1/eV) & (eV) & (meV)  \\
\\[-2.1ex]
\midrule  \addlinespace

$ \rm{Cr_2S_2} $ &  -59.5 & 40.1 & 1.3 & 0.72  & 1.47 & 11.9    \\ 
$ \rm{Mn_2S_2} $ & 0.0 & -297.4 & 2.3 & 0.47  & 1.51  & -28.1   \\  
$ \rm{Mo_2S_2} $ &  -2.7 &  2.7 & 0.5 & 0.87  & 1.34  & 5.4   \\
$ \rm{Cr_2Se_2} $ & -8.8 & -28.3 & 1.3  & 0.53  & 1.34  & -8.4   \\ 
$ \rm{Mn_2Se_2} $ & 0.0  & -438.1 & 2.5  &  0.27  & 1.32 & -35.1  \\  
$ \rm{Fe_2Se_2} $ & -95.9 & -102.1 & 2.7  & 0.61  & 1.38  & -7.0   \\  
$ \rm{Mo_2Se_2} $ &  -1.2 &  1.2 & 0.3 & 0.83  & 1.24  & 9.6  \\
$ \rm{Ti_2Te_2} $ & -29.7 & 29.7 & 1.5  & 1.24 & 0.95 & 6.6   \\ 
$ \rm{V_2Te_2} $ & 0.0 & -6.5 & 0.7  & 0.49 & 1.20 & -6.5   \\ 
$ \rm{Cr_2Te_2} $ & -1.1  & -110.2 & 1.8  & 0.49  & 1.08   & -18.0  \\
$ \rm{Mn_2Te_2} $ & -60.4 & -396.9 & 2.5 & 0.21 & 1.09 & -31.7  \\
$ \rm{Fe_2Te_2} $ & -334.0 & 284.4 & 4.5 & 1.06 & 1.18 & 7.2  \\
$ \rm{Hf_2Te_2} $ & -31.1 & 31.1 & 1.4  & 0.92 & 1.14 & 2.6 \\
\bottomrule
\addlinespace
			\end{tabular}			
			\label{table:magnetic}
	\end{threeparttable}
\end{table}

As most of the M$_2$X$_2$ systems contain partially filled $\textnormal d$
TM atoms in non-magnetic calculation, we can use the simple Stoner model to discuss the
appearance of ferromagnetism in these materials. Based
on this model, the instability of the paramagnetic state towards ferromagnetic ordering is given by 
the criterion $I.D(E_\textnormal F)>1$, where $I$ is the Stoner parameter and $D(E_\textnormal F)$ is the DOS at the $E_\textnormal F$ in 
the nonmagnetic state at the Fermi level.  Within the multi-orbital Hubbard model, the
relationship between the Stoner parameter $I$, Hubbard $U$, and exchange $J$ is given by
$I=(U+6J)/5$~\cite{Stollhoff}. On the basis of the calculated effective Coulomb parameters $U$ 
and $J$ determined by first principles using the cRPA method, $I.D(E_\textnormal F)$ values for ferromagnetic 
compounds are put together in Table\,\ref{table:magnetic}.

For most of the FM M$_2$X$_2$, we observe substantial $D(E_\textnormal F)$ values and subsequently large $I.D(E_\textnormal F)$ values, 
exceeding those found in NM and AFM M$_2$X$_2$ materials. Consequently,  the significant DOS of Ti, Cr, Fe, Mo,
and Hf atoms at the $E_\textnormal F$ in Ti$_2$Te$_2$, Cr$_2$S$_2$, Fe$_2$Te$_2$, Mo$_2$X$_2$ (X=S, Se), and Hf$_2$Te$_2$ respectively can lead to instability in magnetic ordering, which agrees well with our findings from spin-polarized total energy 
calculations and the substantial computed magnetic moments. The spin-polarized density functional 
calculations indicate that the ground states of the V$_2$X$_2$, Co$_2$X$_2$, and Cu$_2$X$_2$ are nonmagnetic because of 
the small $D(E_\textnormal F)$. 

Finally, we derive the magnetic exchange constant $J_H$ for the nearest neighboring M atoms using the total 
energies obtained from DFT calculations. 
The exchange constant $J_H$ is derived from the exchange energy as follows:
\begin{equation}
J_\textnormal H=\frac{E_{\textnormal{AFM}}-E_{\textnormal{FM}}}{2zS_{\textnormal{TM}}^2},
\end{equation} \\
where $E_{\textnormal{AFM}}$ and $E_{\textnormal{FM}}$ are total energy per TM atom for anti-ferromagnetic and ferromagnetic states,
respectively, $z$ is the number of nearest TM neighbors. Here, $z=4$ for single-layer
M$_2$X$_2$. $S_{\textnormal{TM}}$ is the magnetic moment of each TM atom.  Our calculated exchange constants for magnetic 
systems are summarized in Table\,\ref{table:magnetic}. These values of $J_H$ can be employed in model
Hamiltonian calculations to estimate the critical properties of ferromagnetic materials such as
Curie temperature $T_\textnormal C$.

\section{conclusion}\label{sec4}
Employing density functional theory (DFT) calculations, we have predicted a novel family of 2D compounds built around TM under the classification M$_2$X$_2$ (with M symbolizing TM and X representing chalcogen elements like S, Se, and Te).
 These materials can be viewed as a AB stacking of two honeycomb MX layers.  Each MX layer shares a similar structure to graphene but consists of alternating M (sublattice A) and X (sublattice B) atoms arranged in a buckled hexagonal lattice.
Our study encompasses a thorough exploration into the formation energies, dynamical/thermal stabilities, mechanical properties, electronic structures, and magnetic properties of different systems within this compound family. Our computational analyses have unearthed 35 thermodynamically and dynamically stable M$_2$X$_2$ monolayer materials that exhibit remarkable diversity in their electronic and magnetic attributes. These findings are poised to pave the way for the experimental realization of a range of M$_2$X$_2$ structures with elusive and useful properties. 
Specifically, within the assortment of projected compounds, M$_2$X$_2$ (where M=Zn, Cd and X=S, Se, Te) emerge as direct band-gap semiconductors, exhibiting $E_\textnormal g$ ranging from 1.3 to 3.7 eV through hybrid functional calculations. Higher degree of control over the valleys in this new family of semiconductors is also
promising for a new generation of semiconductor based spin qubits. Possibility of employing strain gauge fields similar to those in graphene nail down the valley degree of freedom in this new 2D semiconductors rises hopes for valley non-degenerate spin qubits in this systems. 
On the other hand, the compounds M$_2$X$_2$(M=Ti, Zr, Hf, Tc, Re) behave as zero-gap semiconductors or semimetals when analyzed using standard DFT+PBE calculations. However, the inclusion of spin–orbit coupling  opens a band gap of approximately 0.1 eV in these materials.
This broad range of band gaps suggests a diverse potential for applications in optoelectronic devices, particularly in areas requiring \textit{tunable} electronic properties.
Interestingly, our examination has revealed \textit{intrinsic} magnetic properties in the present class of materials, such as Mn$_2$X$_2$(X=S, Se), Fe$_2$X$_2$(X=Se, Te), and Ti$_2$Se$_2$. The identification of intrinsic magnetism in M$_2$X$_2$ materials not only provides valuable insights into the magnetic behavior of 2D systems but also positions these substances as promising candidates for the advancement of cutting-edge spintronics devices.

\section{Methods}

We have carried out the DFT-based calculations for the M$_2$X$_2$ compounds using projector augmented wave
(PAW) \cite{Blochl} pseudopotentials, as implemented in the
Vienna  \emph{ab initio} simulation package (VASP) \cite{Kresse} within the generalized gradient approximation in PBE parameterization~\cite{Perdew}. 
 In addition, the Heyd–Scuseria–Ernzerhorf
(HSE06) \cite{Heyd,Krukau} functional is used to correct
the underestimated band gaps. The HSE06
functional is applied by mixing 25$\%$ of the exact
Hartree–Fock (HF) exchange potential with 75$\%$ of
PBE exchange and 100$\%$ of PBE correlation energy.
The k-point mesh and the cut-off energies are determined by optimization of total energy to converge energy to 10$^{-8}$ eV/unit-cell.
Therefore, the uniform k-grids of $14\times14\times1$ in the
first Brillouin zone (BZ) is utilized for the self-consistent field
calculation. The optimized kinetic energy cut-off for the wavefunctions is 450 eV for non-SOC, SOC and 550 eV for HSE+SOC calculations. A vacuum layer of 20 $\textnormal \AA$ is
inserted along the non-periodic direction to prevent
the unrealistic interactions between the neighboring
layers. For each system, the Broyden–Fletcher–Goldfarb–Shanno quasi-newton algorithm is
used to relax the internal parameters for the M and X positions to reach sufficiently small forces, as small 
as 10$^{-4}$ eV/$\textnormal \AA$.
Phonon calculations are carried out employing the finite displacement method 3 $\times$ 3 supercell framework, utilizing the VASP-PHONOPY interface. atomic displacements of 0.01 $\textnormal \AA$ are applied to perturb the equilibrium atomic positions. and the resulting forces are computed via DFT+PBE. Phonopy post-processed the VASP-derived force sets to construct the dynamical matrix, enabling precise determination of phonon dispersion relations across high-symmetry paths in the Brillouin zone.
 \emph{Ab initio} molecular dynamics (AIMD) simulations are carried out using the \textit{Nosé}-Hoover thermostat to control the system temperature within a canonical NVT ensemble. A 4 $\times$ 4 $\times$ 1 super-cell is employed here with a time step of 3.0 fs, and a total simulation time of 6500 fs (6.5 ps). 
The second-order elastic constants ($C_{ij}$) are calculated using the stress-strain method with the PBE-GGA approximation. A 16 $\times$ 16 $\times$ 1 Monkhorst-Pack k-point grid and a plane-wave cutoff energy of 550 eV are used for the mechanical calculations. Symmetry-preserving deformations are applied in increments of 0.05 $\%$ strain within the range of 1.5$\%$, and the raw data are processed using the VASPKIT tool. The elastic constants are extracted by polynomial fitting of the stress-strain curves.

We determine the strength of the on-site coulomb $U$ and exchange $J$ interactions
between correlated electrons for magnetic systems within the
cRPA method~\cite{Aryasetiawan,Nomura} as implemented in the SPEX code~\cite{Schindlmayr}. 
In this part, for the ground-state DFT calculations we use the FLEUR
code~\cite{Fluer}, which is based on the full-potential linearized
augmented plane-wave method. Here, the calculations are performed
using cutoff for the wave functions ($\textnormal k_{\textnormal max}$=4 \textnormal a.u.$^{-1}$), and the $16\times16\times1$ k-point grid.
The bands around the $E_{\textnormal F}$, formed
by TM $\textnormal d$-orbitals in M$_2$X$_2$, are chosen as a target correlated subspace. The maximally localized Wannier functions (MLWFs) are constructed using the WANNIER90 library~\cite{Pizzi}. We use a $9\times9\times1$ k-point grid for all structures
in the cRPA calculations.

\subsection{Data availability}
Some more results that support the findings of this study are available in the supplementary material.
Further details if required, are available from the corresponding authors upon request.

\section{Acknowledgements}
S. A. J. was supported by Alexander von Humboldt foundation and EinQuantumNRW

\end{document}